\documentclass[12pt]{article}
\usepackage{amsmath}
\usepackage{cite}
\usepackage{relsize}
\usepackage{amssymb}
\usepackage{graphics}
\usepackage{epsfig}
\usepackage{epstopdf}
\usepackage{verbatim}
\usepackage{color}
\usepackage{slashed}
\usepackage{subcaption}
\usepackage{multirow}
\usepackage{textcomp}
\usepackage{float}
\usepackage[dvipsnames]{xcolor}
\usepackage{mathtools}
\usepackage[toc,page]{appendix}
\usepackage{enumitem}

\begin{document}

\title{\textbf{Electron and Muon $(g-2)_{e,\mu}$ Anomalous Magnetic Moment in $U(1)_{L_e-L_{\mu}}$ Symmetry Model}}

\author {\small{Rishu Verma$_a$\thanks{ rishuvrm274@gmail.com}, Ankush$_{a,b}$\thanks{ ankush.bbau@gmail.com}, and B. C. Chauhan$_a$\thanks{ bcawake@hpcu.ac.in}}}

\date{\textit{$^a$ Department of Physics and Astronomical Science,\\Central University of Himachal Pradesh, Dharamshala 176215, INDIA.}\\
\textit{$^b$ Department of Physics,\\Guru Kashi University, Bathinda(Pb) 151302, \\INDIA.}}

\maketitle

\begin{abstract}
The nature of neutrino (whether Majorana or Dirac) and the origin of neutrino masses are still some of the mysteries to be resolved. Also, the recent results on (g-2)$_{e,\mu}$ measurements deviate from the Standard Model (SM) predictions and motivate us towards new physics beyond the SM. In this work, we propose a model with the minimal field content in the framework of anomaly free extension of Standard Model; i.e. U(1)$_{L_e-L_{\mu}}$ symmetry model. We find this model capable of explaining the low energy neutrino phenomenology and anomalous magnetic moment(g-2)$_{e,\mu}$ of electron and muon, simultaneously. The field content is extended by a SU(2)$_L$ singlet scalar field $\phi$ and three right handed neutrinos N$_R$(R = 1,2,3). Thus, the neutrino masses are generated using the Type-I seesaw mechanism. The extended model leads to the results, which are in consistency with the experimental values of (g-2)$_{e,\mu}$ and also satisfy all the relevant experimental data.
\end{abstract}

\section{Introduction}
The Standard Model (SM) of particle physics is an amazing hypothesis that can satisfactorily explain the interactions of elementary particles and their dynamics. In spite of its enormous success, it falls short in explaining the neutrino mass, matter-antimatter asymmetry, dark matter (DM), anomalous magnetic moment $(g-2)_{e,\mu}$, etc. The solar and atmospheric neutrino oscillation experiments like Super-Kamiokande \cite{SK}, SNO \cite{SNO,SNO1} and KamLand \cite{KAM} confirmed the massive nature and flavor mixing of neutrinos providing the hints for physics beyond standard model(BSM). Despite ongoing experimental advances, the origin of neutrino mass and mixing remains unknown\cite{origin1,origin2}.

\noindent The seesaw mechanism is a reliable theoretical explanation for the smallness of neutrino mass and its mixing\cite{seesaw}. Incorporating right-handed (RH) neutrinos into the Standard Model is a simple technique to explain nonzero neutrino masses. The inclusion of RH neutrinos (RHNs) in the theory therefore makes it possible to formulate a Yukawa term that relates left-handed (LH) and RH neutrino fields to the SM Higgs doublet. The existence of lepton number violation at a certain high energy scale forms the essential basis of the seesaw mechanism. In order to have neutrino mass at sub-eV, the new physics must be at the order of GUT scale, i.e. $10^{16}$ GeV. The different types of seesaw mechanisms therefore, account for the minute non-zero neutrino masses, but they also create a new physics scale that is outside the scope of present and foreseeable accelerator studies. The simplest seesaw mechanism, type-I or canonical \cite{type11,type12,type13, type14, type15, type16} uses only three extra SM singlet RH neutrinos to illustrate the sub-eV masses of neutrinos. The effective neutrino mass matrix in type-I seesaw scenario is given as
\begin{equation}
 m_{\nu} \approx -M_{D}  M_R ^{-1} M_{D}^{T}.\\
 \label{eq:3}
\end{equation}
 where, $M_{R}$ and $M_{D}$ represents the RH and Dirac neutrino mass matrix, respectively.

\noindent On the other hand, the charged leptons' magnetic moments have been projected as effective accuracy tests for the entire SM and quantum electrodynamics (QED). The Fermi National Accelerator Laboratory (FNAL) studies have shown that the experimental value of magnetic moment of muons is not consistent with the SM model prediction, with a 4.2$\sigma$ divergence \cite{FNAL}. Similar to this, the experimental value of the magnetic moment of the electron deviates from the standard model prediction by 1$\sigma$ and 2.4$\sigma$, based on measurements of the rubidium atom and the cesium atom, respectively \cite{rubidium,cesium}. 

\noindent  All these conflicts make it more likely towards the repercussions that go beyond the SM (BSM). Gauged lepton flavor models, such as $U(1)_{L_e-L_\mu}$, not only address the issue of light neutrino mass but also naturally account for the muon (g-2) anomaly within a minimal setup \cite{gmodel, gmodel1, gmodel2, gmodel15, gmodel16}. In this work, we considered the extension of SM with extra gauged $U(1)_{L_e-L_\mu}$ symmetry in order to explain neutrino mass, mixing and electron and muon anomalous magnetic moment $(g-2)_{e,\mu}$ in a common framework. Moreover, the gauged symmetry $U(1)_{L_e-L_\mu}$ is characterized by being anomaly free \cite{gmodel3, gmodel4}. 
 Recent investigations on $U(1)_{L_e-L_\mu}$ models concerning the electron and muon (g-2) anomalies  employing extended inverse seesaw and inverse see saw(2,3) mechanisms are documented in \cite{rk, rv}. However, we have put forward a simplistic model that employs the type-I seesaw mechanism to produce nonzero sub-eV neutrino masses \cite{type11, type12, type13, type14, type15, type16}. The SM is extended by three right handed neutrinos $N_i (i = 1,2,3)$ as required for type-I seesaw framework. Within the framework of the gauged $U(1)_{L_{\mu}-L_\tau}$ model, researchers have also included a type-I seesaw mechanism to explain the origin of light neutrino masses. Two extra Higgs doublets have been added to this extended model, and studies of the muon (g-2) anomaly have also been carried out in this framework \cite{debo}. However, in our $U(1)_{L_e-L_\mu}$ model, the field content is expanded solely by the inclusion of an additional scalar $SU(2)_L$ singlet field, denoted as $\phi$, which serves to address the electron and muon anomalous magnetic moment, simultaneously while maintaining the model's minimal nature. The latest data and constraints from various experimental observations are used for numerical calculations.

\noindent This paper is organized as follows: Section \ref{sec:2} contains the complete description of the model components, Lagrangian and generation of neutrino mass matrices. In Section \ref{sec:3}, anomalous magnetic moment of muon and electron are explained. Numerical analysis and final results are discussed in Section \ref{sec:4}. Finally, in Section \ref{sec:5}, the conclusions are given.
  
  \section{The Model}\label{sec:2}

The model is a minimal extension of the SM based on gauged $U(1)_{L_e-L_\mu}$ symmetry, which is an anomaly free symmetry. We incorporated three RHNs, $N_{i}(i = e,\mu,\tau)$ and a scalar singlet $\phi$ in addition to SM particles. The charge assignments of $N_{i}$ and $\phi$ under $U(1)_{L_e-L_\mu}$ symmetry is (-1,1,0) and 1, respectively. The light neutrino masses are obtained by using type-I seesaw mechanism. The fermionic and scalar field content along with respective charge assignments are shown in Table \ref{table1}.

\begin{center}
\begin{table}[h]
\centering
\begin{tabular}{cccccccccccc}
 Symmetry & $\bar{L}_{e}$ & $\bar{L}_{\mu}$ & $\bar{L}_{\tau}$ & $e_{R}$ & $\mu_{R}$ & $\tau_{R}$ & $N_{e}$ & $N_{\mu}$ &  $N_{\tau}$ & $H$  & $\phi$ \\
 \hline
$SU(2)_L$    &     2  & 2 & 2 & 1  & 1 & 1   & 1   &   1     &    1        &   2        &    1    \\

\hline
$U(1)_{L_{e}-L{\mu}}$ & 1  & -1 & 0  &   -1    & 1  &  0  & -1 & 1 &  0 & 0 & 1  \\
\hline

\end{tabular}
 \caption{Fermion and scalar field content and respective charge assignments under $SU(2)_L \times U(1)_{L_{e}-L{\mu}}$ model.}
 \label{table1}
\end{table}
\end{center}

\noindent The relevant Lagrangian for the proposed model is
\begin{eqnarray}
\nonumber
\mathcal{L} \subseteq &&\bar{N}_e\iota\gamma^{\mu}D_{\mu}N_e +\bar{N}_{\mu}\iota\gamma^{\mu}D_{\mu}N_{\mu}-\frac{M_{e\mu}}{2}N_{e}N_{\mu}-\frac{M_{\tau\tau}}{2}N_{\tau}N_{\tau}-y^1_{\phi}N_{\mu}N_{\tau}\phi^{\ast}- y^2_{\phi}N_{e}N_{\tau}\phi-\\
\nonumber
&&y_{e}^{\nu}\Bar{L}_{e}N_{e} \Tilde{H} - y_{\mu}^{\nu}\Bar{L}_{\mu}N_{\mu} \Tilde{H} -
y_{\tau}^{\nu}\Bar{L}_{\tau}N_{\tau} \Tilde{H}+y_{e}\Bar{L}_{e}e_{R}H + y_{\mu}\Bar{L}_{\mu}\mu_{R}H + y_{\tau}\Bar{L}{_\tau}\tau_{R}H + h.c.,
 \label{eq:4}
\end{eqnarray}

\noindent where $\Tilde{H}=i\tau_{3}H$ and $y_q(q = e,\mu,\tau)$, $y_{i}^{\nu} (i =e,\mu,\tau )$, $y_{\phi}$ are Yukawa coupling constants. The neutral component of Higgs is responsible for breaking the electroweak symmetry and the scalar singlet $\phi$ is breaking the $ U(1)_{L_{e}-L{\mu}}$ symmetry after acquiring following non-zero vacuum expectation values
\begin{center}
$\langle H \rangle = v_{H}$ and 
$\langle \phi \rangle = v_{\phi}$.
\end{center}
\noindent The covariant derivatives of the RHN are\\
\begin{equation}
  \slashed{D}N_e   = (\slashed{\partial} + ixg_{e\mu} \slashed{ Z}_{e\mu})N_e,
  \end{equation}
  \begin{equation}
        \slashed{D}N_{\mu}   = (\slashed{\partial} + ixg_{e\mu}\slashed{ Z}_{e\mu})N_{\mu}.
\end{equation}
\noindent As a consequence, we have obtained a diagonal charged lepton mass matrix as\begin{equation}
m_{l}= Diag(y_{e},y_{\mu},y_{\tau})v_{h}.\\
\end{equation}   
 
\noindent The other mass matrices which we have obtained are shown as below \\
\begin{equation}
m_{D} =\begin{pmatrix}
a & 0 &0 \\
0 & b &0\\
0 & 0 &h\\
 \end{pmatrix}, M_R = \begin{pmatrix}
0 & t & y \\
t & 0 & z  \\
y & z & l\\
 \end{pmatrix},
\end{equation}

\noindent where $a=y_{1}^{\nu}v_{H}, b = y_{2}^{\nu}v_{H},  h = y_{3}^{\nu}v_{H}, t = M_{e\mu}, y =  y_{\phi}^2v_{\phi},z =  y_{\phi}^1v_{\phi},l = M_{\tau\tau}$. Within Type-I seesaw mechanism, the above matrices lead to the light neutrino mass matrix as follow\\
\begin{equation}
 m_{\nu} = -m_dM_R^{-1}m_d^T\\
= \begin{pmatrix}
\frac{t^2}{b}+ \frac{y^2}{h}& \frac{yz}{h} & \frac{ly}{h}+\frac{tz}{b} \\
 \frac{yz}{h} &\frac{t^2}{a} +\frac{z^2}{h} & \frac{ty}{a}+\frac{lz}{h} \\
\frac{ly}{h}+\frac{tz}{b} &\frac{ty}{a}+\frac{lz}{h} &\frac{l^2}{h} +\frac{y^2}{a}+\frac{z^2}{b} \\
\end{pmatrix}.
\label{eq:12}
\end{equation}

\section{ Muon and Electron (g-2) Anomaly}\label{sec:3}
\begin{figure}[H]
	\begin{center}
			{\epsfig{file=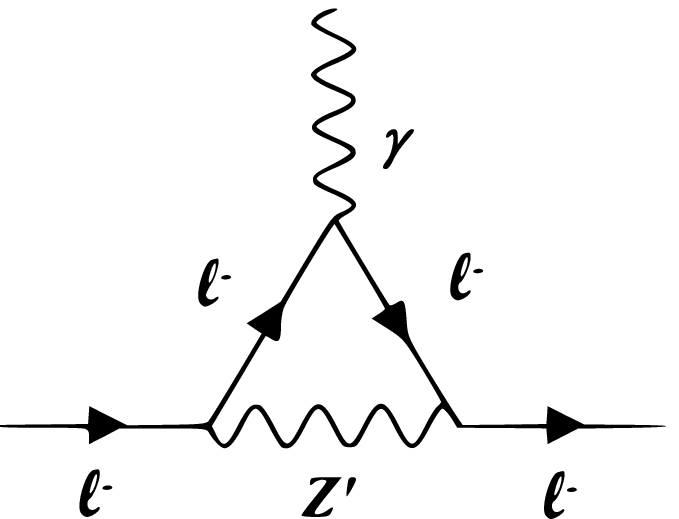,height=4.0cm,width=7.0cm}}
		\end{center}
\caption{\label{fig1}Feynman diagram for electron and muon ($l=e,\mu$) g-2 with mediator Z$'$ gauge boson.}
\end{figure}

\subsection{Muon (g-2) Anomaly}
For any elementary particle with charge `$q$' and spin $\Vec{S}$, the magnetic moment $\Vec{\mu}$ is given as
\begin{equation}
    \Vec{\mu} = g\frac{q}{2m}\Vec{S},
\end{equation}
\noindent where $m$ represents the mass of the particle and $g$ is the gyromagnetic ratio. Using a quantum mechanical model, Dirac predicted the value of $g$ for the electron in 1928 that for any spin-1/2 particle equals 2 \cite{Dirac}. The radiative adjustments, however, tend to raise its value from 2 at higher orders. The term anomalous magnetic moment ($a$) specifically refers to the fractional divergence of $g$ from Dirac's prediction. The anomalous magnetic moment for muons is given by the formula $a_{\mu} = (g-2)/2$. The value of $a_{\mu}$ is predicted by the most recent muon (g-2) experiment results from the Fermi National Accelerator Laboratory (FNAL)\cite{FNAL} as
\begin{equation}
    a_{\mu}^{FNAL} = 116592040(54)\times10^{-11}.
    \label{eq:21}
\end{equation}
On the other hand, the theoretical predictions of SM state that \cite{SM}
\begin{equation}
    a_{\mu}^{SM} = 116591810(43)\times10^{-11}.
    \label{eq:22}
\end{equation}
This 4.2$\sigma$ significant discrepancy  $ \Delta a_{\mu} ( a_{\mu}^{FNAL}-a_{\mu}^{SM}$) questions the SM and points to novel physics that goes beyond it. Using Eq.(\ref{eq:21}) and Eq.(\ref{eq:22}), we obtain
\begin{equation}
    \Delta a_{\mu} = 251(59)\times10^{-11}.
\end{equation}
\noindent
The anomalous magnetic moment of the electron and muon are important quantities that have been measured to high precision and provide important tests of the predictions of the Standard Model of particle physics. The literature contains a number of models that have discussed $(g-2)_{\mu}$ anomaly \cite{g1,g2,g3,g4,g5,g6,g7,g8,g9,g10,g11}. Since we have extended our model by $U(1)_D$ symmetry, the $Z'$ can contribute to $(g-2)_\mu$ anomaly if the its mass is in the range 10-100 MeV. Through the interaction of muon with $Z'$ can provide substantial rectification to $g-2$. The neutral current interaction that contributes to $(g-2)_{\mu}$'s calculation is provided as
\begin{equation}
    \mathcal{L} = - g'\bar\mu Z' \gamma^{\mu} \mu,
\end{equation}
\noindent where $g'$ represents the corresponding coupling constant.The analytical one loop contribution of $Z'$ can be expressed as\cite{analytical1,analytical2}
\begin{equation}
    \Delta a_{\mu}^{Z'} = \frac{g'^2}{8\pi^2}\int_{0}^{1}dx\frac{2m_{\mu}^2 x^2(1-x)}{x^2m_{\mu}^2+(1-x)M_{Z'}^2},
\end{equation}

where $m_{\mu}$ is the mass of muon and $M_{Z'}$ is the gauge boson mass.

\subsection{Electron (g-2) Anomaly}
Recent experiments have not been able to precisely detect the magnetic moment of the electron, unlike the magnetic moment of the muon. The measurement of the Rubidium atom determined the electron's magnetic moment as \cite{rubidium}
\begin{equation}
    (\Delta a_{e})_{Rb} = 48(30)\times10^{-14}
\end{equation}

\noindent with $1\sigma$ discrepancy over SM, while the Cesium atom experiment produces the magnetic moment of electron as
\begin{equation}
    (\Delta a_{e})_{Cs} = -87(30)\times10^{-14}
\end{equation}
\noindent with $2.4\sigma$ discrepancy over SM \cite{cesium}. 
It is obvious that there is a lot of uncertainty about signs.

\noindent The current experimental values for $\Delta a_{e}$ and $\Delta a_{\mu}$ agree with the theoretical predictions to within a few parts in a billion and show a significant discrepancy, respectively, which could be an indication of new physics. In order to explain these discrepancies in electron and muon anomalous magnetic moment, anomaly free gauge symmetry $U(1)_{L_{e}-L{\mu}}$ is used. The extra gauge boson $Z'$ (in MeV range) obtained from $U(1)_{L_{e}-L{\mu}}$ symmetry breaking effectively contributes to $\Delta a_{\mu}$ and $\Delta a_{e}$ . The neutral current interaction which gives the contribution to the calculation of $(g-2)_e$ is given as
\begin{equation}
    \mathcal{L} = - g'\bar e Z' \gamma^{e} e.
\end{equation}
  The one loop contribution of $Z'$ to the magnetic moment of electron is given as \cite{electron}
\begin{equation}
    \Delta a_{e}^{Z'} = \frac{g'^2}{8\pi^2}\int_{0}^{1}dx\frac{2m_{e}^2 x^2(1-x)}{x^2m_{e}^2+(1-x)M_{Z'}^2},
\end{equation}
\noindent where $m_{e}$ is mass of electron. The contributing Feynman diagram is shown in Fig. \ref{fig1}.  

\section{Numerical Analysis and Results}\label{sec:4}
This section contains a numerical estimate of the model's viability using the data on neutrino oscillations and constraints on $\Delta a_{e,\mu}$. It is evident from the light neutrino mass matrix $M_{\nu}$ obtained by employing type-I seesaw in Eq.(\ref{eq:12}), that we have eight unknown parameters. These model parameters are evaluated numerically using the neutrino oscillation data constraints as presented in Table \ref{table4}.

\begin{table}[h]
\begin{center}
\begin{tabular}{c|c|c}
\hline \hline 
Parameter & Best fit $\pm$ \( 1 \sigma \) range & \( 3 \sigma \) range  \\
\hline \multicolumn{2}{c} { Normal neutrino mass ordering \( \left(m_{1}<m_{2}<m_{3}\right) \)} \\
\hline \( \sin ^{2} \theta_{12} \) & $0.304^{+0.013}_{-0.012}$ & \( 0.269-0.343 \)  \\
\( \sin ^{2} \theta_{13} \) & $0.02221^{+0.00068}_{-0.00062}$ & \( 0.02034-0.02420 \) \\
\( \sin ^{2} \theta_{23} \) & $0.570^{+0.018}_{-0.024}$ & \( 0.407-0.618 \)  \\
\( \Delta m_{21}^{2}\left[10^{-5} \mathrm{eV}^{2}\right] \) & $7.42^{+0.21}_{-0.20}$& \( 6.82-8.04 \) \\
\( \Delta m_{31}^{2}\left[10^{-3} \mathrm{eV}^{2}\right] \) & $+2.541^{+0.028}_{-0.027}$ & \( +2.431-+2.598 \) \\
\hline \multicolumn{2}{c} { Inverted neutrino mass ordering \( \left(m_{3}<m_{1}<m_{2}\right) \)} \\
\hline \( \sin ^{2} \theta_{12} \) & $0.304^{+0.013}_{-0.012}$ & \( 0.269-0.343 \)\\
\( \sin ^{2} \theta_{13} \) & $0.02240^{+0.00062}_{-0.00062}$ & \( 0.02053-0.02436 \) \\
\( \sin ^{2} \theta_{23} \) & $0.575^{+0.017}_{-0.021}$& \( 0.411-0.621 \) \\
\( \Delta m_{21}^{2}\left[10^{-5} \mathrm{eV}^{2}\right] \) & $7.42^{+0.21}_{-0.20}$ & \( 6.82-8.04 \) \\
\( \Delta m_{32}^{2}\left[10^{-3} \mathrm{eV}^{2}\right] \) & $-2.497^{+0.028}_{-0.028}$ & \( -2.583--2.412 \)  \\
\hline \hline
\end{tabular}
\end{center}
\caption{Current best fit neutrino oscillations experimental data NuFIT 5.0 used in the numerical analysis\cite{data}.}
\label{table4}
\end{table}

\noindent The probability of a neutrino changing its flavor is determined by the Pontecorvo-Maki-Nakagawa-Sakata (PMNS) matrix, which is also known as the neutrino mixing matrix. The PMNS matrix describes how the three flavor states of neutrinos (e, $\mu$ and $\tau$) are related to the three mass states of neutrinos ($m_1$, $m_2$, and $m_3$). The PMNS matrix has three mixing angles ($\theta_{12}$, $\theta_{23}$ and $\theta_{13}$) and one phase ($\delta$), which are currently known with varying degrees of precision. The phase manages the interference between the various mass states, whilst the mixing angles govern the degree of mixing between the flavour and mass states. We have six neutrino oscillation parameters, including the two independent mass squared differences, $\Delta m_{32}^{2}$ and $\Delta m_{21}^{2}$. In charged lepton basis, the light neutrino mass matrix can be written as
\begin{equation}
M_{\nu}=UM_{d}U^T,
\end{equation}
\noindent where $U$ is PMNS $(U_{PMNS})$ neutrino mixing matrix and $M_{d}$ is diagonal mass matrix containing mass eigenvalues of neutrinos $diag(m_{1}, m_{2}, m_{3})$.  $U_{PMNS}=V.P$ 
where $ P$ is diagonal phase matrix  $diag(1,e^{i\alpha},e^{i(\beta+\delta)})$, in which $\alpha$, $\beta$ are Majorana type $CP$ violating phases. According to  PDG representation, $V$ is written as
  \begin{equation}
   \begin{pmatrix}
c_{12} c_{13} & s_{12} c_{13} &  s_{13} e^{-i\delta} \\
-s_{12} c_{23} - c_{12} s_{23} s_{13} e^{i\delta} & c_{12} c_{23} - s_{12} s_{23} s_{13} e^{i\delta} &  s_{23} c_{13} \\
s_{12} s_{23} - c_{12} c_{23} s_{13} e^{i\delta} & -c_{12} s_{23} -s_{12} c_{23} s_{13} e^{i\delta} &  c_{23} c_{13} \\
 \end{pmatrix},
   \end{equation}
\noindent where $\delta$ is Dirac $CP$ violating phase. The oscillation parameters $sin^2 \theta_{13}$, $sin^2 \theta_{23}$ and $sin^2 \theta_{12}$ are defined as\\

$sin^2 \theta_{13} = |U_{e3}|^2$, \hspace{1cm} $sin^2 \theta_{23} = \frac{|U_{\mu 3}|^2}{1-|U_{e 3}|^2}$\hspace{1cm} and \hspace{1cm} $sin^2 \theta_{12} = \frac{|U_{e2}|^2}{1-|U_{e 3}|^2}$,\\

\noindent where $U_{\alpha i}$($\alpha$ = e,$\mu$,$\tau$, and i = 1,2,3) represents the $U_{PMNS}$ matrix elements. To check the viability of the model, we used the neutrino oscillation data as given in Table \ref{table4}. Further, we found the parameter space of the model satisfying the neutrino oscillation data. 

\begin{figure}[h]
 \begin{subfigure}[b]{0.4\textwidth}
 \includegraphics[scale=.7]{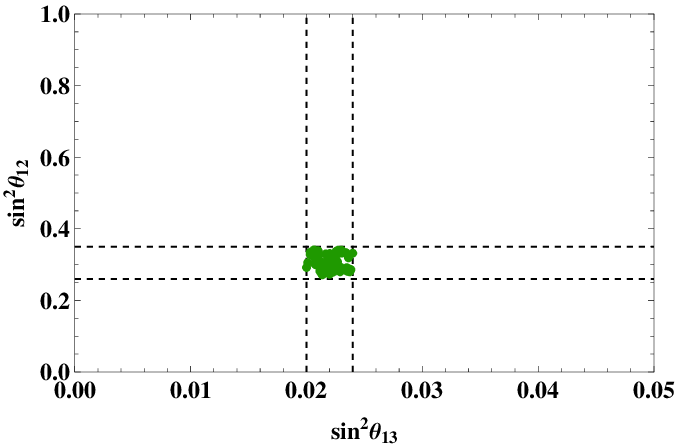}
 \caption{}
 \end{subfigure}
 ~\qquad
 \hspace{.5cm}
 \begin{subfigure}[b]{0.4\textwidth}
 \includegraphics[scale=.7]{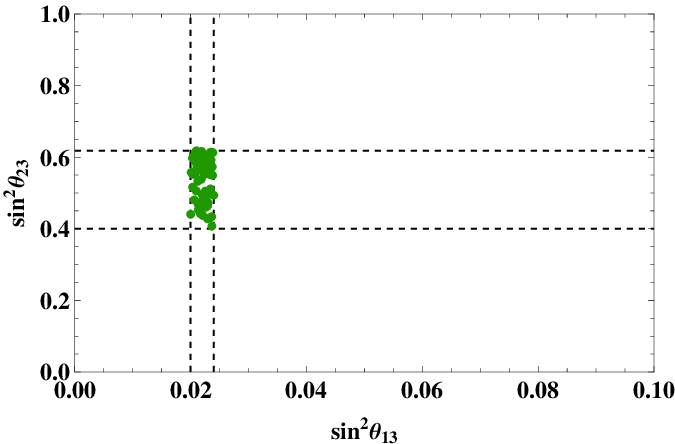}
 \caption{}
 \end{subfigure}
 
 \vspace{1cm}
 \begin{center}
\hspace{-2.5cm}
 \begin{subfigure}[b]{0.4\textwidth}
 \includegraphics[scale=.7]{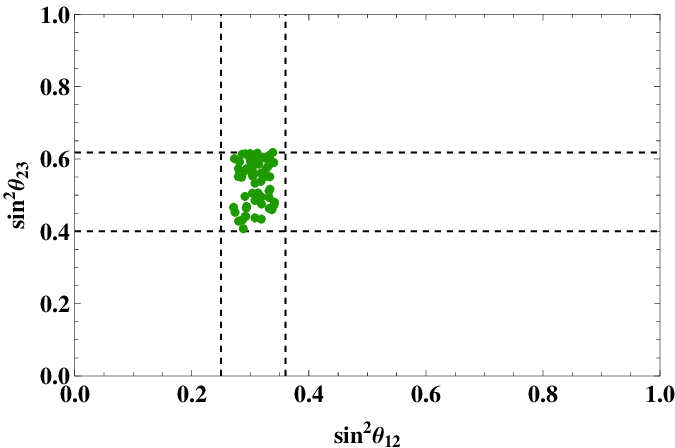}
 \caption{}
 \end{subfigure}
 \end{center}
 \caption{Plots showing correlation of (a)$sin^2 \theta_{12}$ vs $sin^2 \theta_{13}$ , (b) $sin^2 \theta_{23}$ vs $sin^2 \theta_{13}$ and (c) $sin^2 \theta_{23}$ vs $sin^2 \theta_{12}$.}
 \label{Fig2}
 \end{figure}

\newpage
\begin{figure}[h]		\includegraphics[scale=0.7]{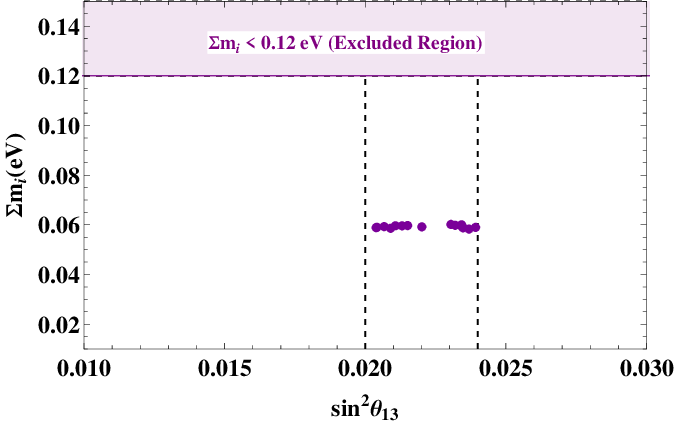}
\includegraphics[scale=0.7]{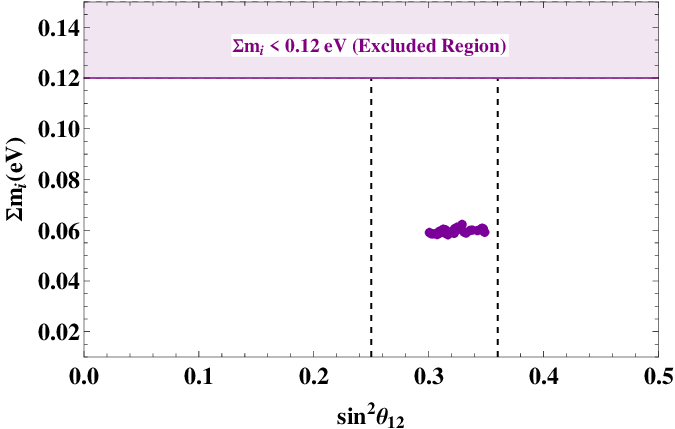}
\caption{ Plots showing correlation of $sin^2\theta_{13}$ (left) and  $sin^2\theta_{12}$ (right) with $\Sigma m_{\nu_{i}}$.}
\label{Fig3}
\end{figure}

\begin{center}
\begin{figure}[h]		
\hspace{3cm}\includegraphics[scale=0.7]{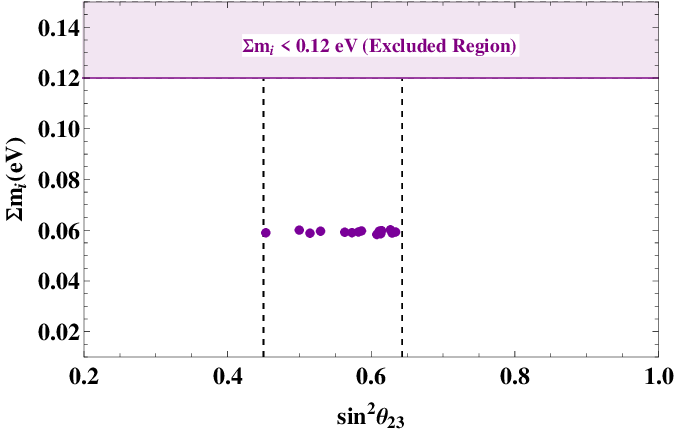}
\caption{ Plot showing correlation of $sin^2\theta_{23}$ with $\Sigma m_{\nu_{i}}$.}
\label{Fig4}
\end{figure}
\end{center}

\newpage
\begin{figure}[h]		\includegraphics[scale=0.7]{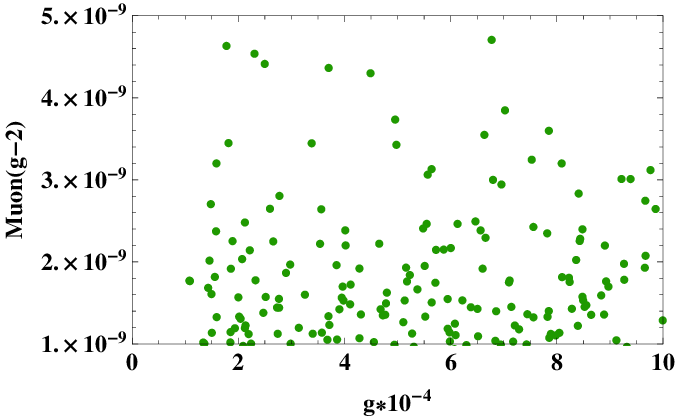}
\includegraphics[scale=0.7]{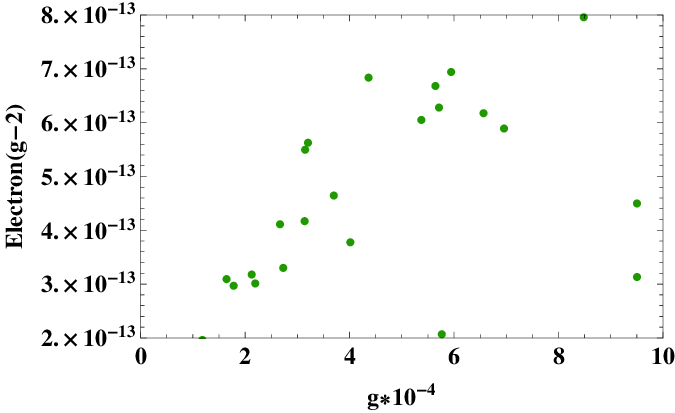}
\caption{ Variation of electron and muon anomalous magnetic moment with gauge coupling constant $g'$.}
\label{Fig5}
\end{figure}

\begin{center}
\begin{figure}[h]		
\hspace{3cm}\includegraphics[scale=0.6]{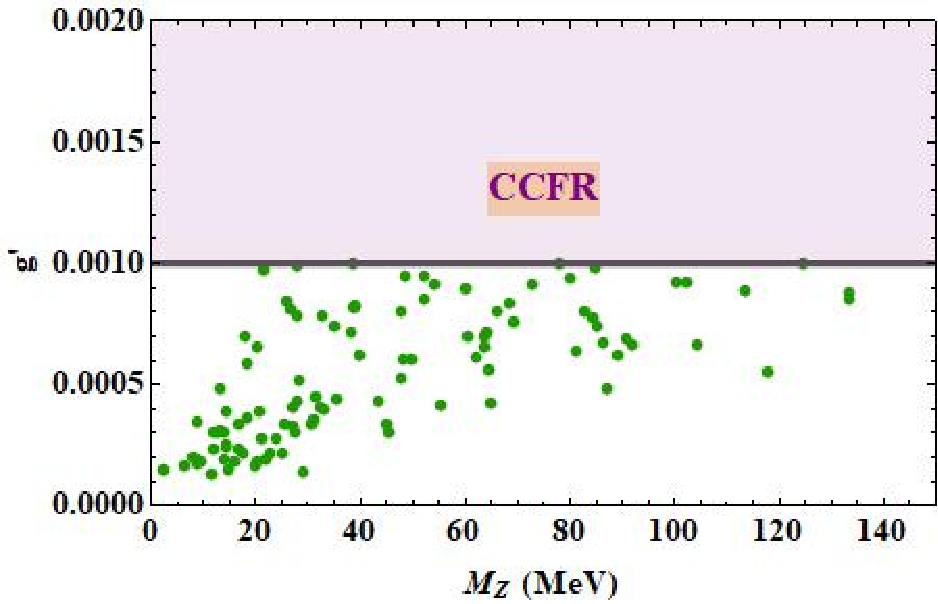}
\caption{  Variation of gauge coupling constant $g'$ with gauge boson mass $M_Z$.}
\label{Fig6}
\end{figure}
\end{center}
\noindent The Fig. \ref{Fig2}(a) represents the correlation of  $sin^2 \theta_{12}$ with $sin^2 \theta_{13}$. Fig. \ref{Fig2}(b) gives the correlation of $sin^2 \theta_{23}$ with $sin^2 \theta_{13}$ and Fig. \ref{Fig2}(c) represents the variation of $sin^2 \theta_{23}$ with $sin^2\theta_{12}$, for normal ordering(NO). It can be seen from Fig. \ref{Fig2}, that the neutrino oscillation data on neutrino masses and mixing is clearly satisfied by the model.

\noindent In Fig. \ref{Fig3}(left), we have shown the correlation of  $sin^2\theta_{13}$  with sum of active neutrino masses, $\Sigma m_{\nu_{i}}$ and  Fig. \ref{Fig3}(right) represents the correlation of $sin^2\theta_{12}$ with $\Sigma m_{\nu_{i}}$, where $m_{\nu_{i}}$ = ($m_{\nu_{1}}$ + $m_{\nu_{2}}$ + $m_{\nu_{3}}$). Fig. \ref{Fig4} represents the correlation of $sin^2\theta_{23}$ with $\Sigma m_{\nu_{i}}$. These plots confirms that the current neutrino oscillation data is satisfied by the model.

\noindent Fig. \ref{Fig5} represents the variation of gauge coupling constant $g'$ with the anomalous magnetic moment(g-2). The left penal represents the variation of gauge coupling constant $g'$ with the muon anomalous magnetic moment $\Delta a_{\mu}$. The right penal shows the variation of  gauge coupling constant $g'$ with the electron anomalous magnetic moment $\Delta a_{e}$. The contribution of extra $Z'$
gauge boson towards anomalous magnetic moment of electron and muon resolves the observed discrepancy in SM. 

\noindent Fig. \ref{Fig6} shows the variation of gauge coupling constant $g'$ with the gauge boson $Z'$ mass $M_{Z'}$ adhering to the CCFR limit. The gauge coupling constant $g'$ is clearly in accordance with the CCFR limit, indicating strong consistency between model's predictions and experimental observations.

\section{Conclusions}\label{sec:5}
In this paper, we have offered a comprehensive analysis of neutrino phenomenology and a combined analysis of anomalous electron and muon magnetic moments $(g-2)_{e,\mu}$. Here, we have employed the simplest and viable type-I scenario to obtain the sub-eV neutrino masses. In addition to the SM particles, three extra RHNs are included in the model. Gauged lepton flavour models, such as $U(1)_{L_e-L_{\mu}}$, provide a natural origin of the muon (g-2) in a relatively minimal setup while also addressing the issue of light neutrino mass. In order for the model to explain the anomalous magnetic moment of the electron and muon $(g-2)_{e,\mu}$ concurrently, the anomaly free $U(1)_{L_e-L_{\mu}}$ gauge symmetry is employed to obtain an additional $Z'$ gauge boson in the MeV range. This MeV range gauge boson is used to explain the anomalous magnetic moment of electron and muon, simultaneously. It is to be noted that the model successfully explains the neutrino phenomenology. The acceptance for normal ordering neutrino occilation data is shown in all the plots associated with the sum of active neutrino masses ($\Sigma m_{\nu_{i}}$). Also, the observed deviation in the Standard Model's anomalous magnetic moment of electrons and muons is resolved by incorporating the contribution of an additional $Z'$ gauge boson in the model. This inclusion effectively bridges the gap between theoretical predictions and experimental measurements.

\noindent In summary, we developed a model formalism which could simultaneously explain the correct neutrino phenomenology and anomalous magnetic moment of electron and muon$(g-2)_{e,\mu}$ and open a window of physics beyond SM. Forthcoming experiments will be able to test the model's predictions, shed light on the nature of neutrinos and dark matter, and offer valuable insights into the fundamental laws of universe.

\vspace{1cm}
\hspace{-.4cm}\textbf{\Large{Acknowledgments}}
 \vspace{.3cm}\\
  B. C. Chauhan is thankful to the Inter University Centre for Astronomy and Astrophysics (IUCAA) for providing necessary facilities during the completion of this work. Ankush acknowledges the financial support provided by the University Grants Commission, Government of India vide registration number 201819-NFO-2018-19-OBCHIM-75542.

\end{document}